\documentclass[preprint,showpacs,preprintnumbers,amsmath,amssymb,nofootinbib]{revtex4}
\pdfoutput=1
\usepackage{booktabs}
\usepackage{mathrsfs}
\usepackage{epsfig}
\usepackage{graphicx}
\usepackage{dcolumn}
\usepackage{bm}
\usepackage{amsmath}
\usepackage{slashed}
\usepackage{multirow}
\usepackage{subfigure}
\usepackage{color}

\let\jnfont=\rm
\def\NPB#1,{{\jnfont Nucl.\ Phys.\ B }{\bf #1},}
\def\PLB#1,{{\jnfont Phys.\ Lett.\ B }{\bf #1},}
\def\EPJC#1,{{\jnfont Eur.\ Phys.\ Jour.\ C }{\bf #1},}
\def\PRD#1,{{\jnfont Phys.\ Rev.\ D }{\bf #1},}
\def\PRL#1,{{\jnfont Phys.\ Rev.\ Lett.\ }{\bf #1},}
\def\MPLA#1,{{\jnfont Mod.\ Phys.\ Lett.\ A }{\bf #1},}
\def\JPG#1,{{\jnfont J.\ Phys.\ G}{\bf #1},}
\def\CTP#1,{{\jnfont Commun.\ Theor.\ Phys.\ }{\bf #1},}
\def\ZPC#1,{{\jnfont Z.\ Phys.\ C }{\bf #1},}
\def\JHEP#1,{{\jnfont JHEP \ }{\bf #1},}
\def\Rv{\not{\hbox{\kern-1pt $R$}}}
\def\p{\not{\hbox{\kern-3pt $p$}}}

\newcommand{\bea}{\begin{eqnarray}}
\newcommand{\eea}{\end{eqnarray}}

\newcommand{\bcen}{\begin{center}}
\newcommand{\ecen}{\end{center}}

\newcommand{\beq}{\begin{eqnarray}}
\newcommand{\eeq}{\end{eqnarray}}

\def\t1{\tilde{t_1}}

\begin{document}

\preprint{
\begin{minipage}[b]{0.75\linewidth}
\begin{flushright}
CTPU-16-40 \\
 \end{flushright}
\end{minipage}
}

\title{Leptonic mono-top from single stop production at the LHC}
\author{ Guang Hua Duan$^{1}$}
\author{ Ken-ichi Hikasa$^{2}$}
\author{ Lei Wu$^{3,4}$}
\author{ Jin Min Yang$^{1,2}$}
\author{ Mengchao Zhang$^5$}

\affiliation{
 $^1$ CAS Key Laboratory of Theoretical Physics, Institute of Theoretical Physics, Chinese Academy of Sciences, Beijing 100190, China\\
 $^2$ Department of Physics, Tohoku University, Sendai 980-8578, Japan\\
 $^3$ ARC Centre of Excellence for Particle Physics at the Terascale, School of Physics, The University of Sydney, NSW 2006, Australia\\
 $^4$ Department of Physics and Institute of Theoretical Physics, Nanjing Normal University, Nanjing, Jiangsu 210023, China\\
 $^5$ Center for Theoretical Physics and Universe, Institute for Basic Science (IBS),
Daejeon 34051, Korea.
 }%


\begin{abstract}
Top squark (stop) can be produced via QCD interaction but also the electroweak interaction at the LHC. In this paper, we investigate the observability of the associated production of stop and chargino, $pp \to \tilde{t}_1\tilde{\chi}^-_1$, in compressed electroweakino scenarios at 14 TeV LHC. Due to small mass splitting between the lightest neutralino ($\tilde{\chi}^0_1$) and chargino ($\tilde{\chi}^-_1$), the single stop production can give the mono-top signature through the stop decay $\tilde{t}_1 \to t \tilde{\chi}^0_1$. We analyze the leptonic mono-top channel of the single stop production and propose a lab-frame observable $\cos\theta_{b\ell}$ to reduce the SM backgrounds. We find that such leptonic mono-top events from the single stop production can be probed at $2\sigma$ level at the HL-LHC if $m_{\tilde{t}_1}<760$ GeV and $m_{\tilde{\chi}^0_1}<150$ GeV. Given a discovery of the stop and a measurement of the single stop production cross
section, the stop mixing angle can also be determined from the single stop production at the HL-LHC.

\end{abstract}

\maketitle

\section{Introduction}
After the discovery of the Higgs boson in 2012 \cite{higgs-atlas,higgs-cms}, the fundamental mechanism of
stabilizing the electroweak scale has become an urgent topic. Weak scale supersymmetry (SUSY) is one of the
most promising candidates for addressing such an longstanding theoretical issue.
SUSY predicts a plethora of supersymmetric particles, among which the top-squarks (stops) play an important
role in cancelling the quadratic divergence in the Higgs boson mass. Naturalness (absence of fine-tuning
in the Higgs boson mass) requires stop masses to be below 1 TeV in the MSSM \cite{bg}. Therefore, the search
for light stops is a sensitive probe of the naturalness in SUSY \cite{nsusy-1,nsusy-2,nsusy-3,nsusy-4,nsusy-5,Polarization,nsusy-6,nsusy-7,nsusy-8,nsusy-9,dutta,nsusy-10,nsusy-11,nsusy-12,Backovic:2015rwa,nsusy-13,nsusy-14,nsusy-15,Goncalves:2016tft,Goncalves:2016nil}.

So far, ATLAS and CMS collaborations have performed extensive searches for stops at the LHC Run-1
and Run-2. The current search strategies are specialized for different kinematical regions.
For example, when $m_{\tilde{t}_1} \gg m_{\tilde{\chi}^0_1} + m_t$, the top quarks from stop decays are usually energetic. With the endpoint observables, such as $M_{T2}$ \cite{Lester:1999tx,Barr:2003rg}, the stop pair can be discriminated from the $t\bar{t}$ background. But in the compressed regions, for example $m_{\tilde{t}_1} \approx m_{\tilde{\chi}^0_1} + m_t$, the decay products of the stop are very soft. In this region, the stop is searched for by using the monojet signature \cite{drees-1,drees-2,qishu,litong,monojet-like-1,monojet-like-2,monojet-like-3}.
Based on recent Run-2 ($\sim 15$ fb$^{-1}$) dataset, the stop mass has been excluded up
to $\sim$ 1 TeV in simplified models
\cite{Aaboud:2016tnv,CMS-PAS-SUS-16-016,nsusy-13-1,nsusy-13-2,nsusy-13-3,nsusy-13-4,nsusy-13-5,nsusy-13-6}.

\begin{figure}[ht!]
   \centering
  \includegraphics[width=4in]{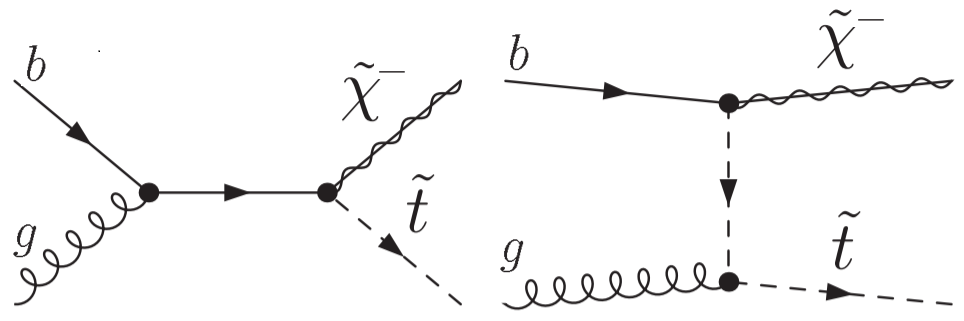}
  \caption{ Feynman diagrams for the associated production of the stop and chargino at the LHC.}
  \label{fig:feyn}
\end{figure}
Like the top quark, stops can be produced in pair, but also can be singly produced via the electroweak interaction, such as the associated production of the stop and chargino, $bg \to \tilde{t}_1 \tilde{\chi}^{-}_{1}$ (c.f. Fig.~\ref{fig:feyn}) \cite{single-stop-1,single-stop-2,wu}. When the stop and chargino are not heavy or the chargino is much lighter than the stop \cite{single-stop-1}, the single stop production can
have a sizable cross section at the LHC. Although the single stop may not be a good discovery channel as the stop
pair production, the study of single stop is meaningful because it can serve as a complementary channel to probe the
electroweak properties of the stop \cite{single-stop-2,wu}.

In this work, we explore the feasibility of probing the single stop production
process $pp \to \tilde{t}_1 ( \to t \tilde{\chi}^0_1) \tilde{\chi}^-_1 + X$ in a compressed
SUSY scenario, where the chargino $\tilde{\chi}_1^\pm$ is almost degenerate with the lightest
neutralino $\tilde{\chi}^0_1$. Such a spectrum is motivated by natural SUSY \cite{nsusy-1,nsusy-2} or
the well-tempered neutralino frameworks \cite{well-tempered}. Due to the small mass splitting between
$\tilde{\chi}^\pm_1$ and $\tilde{\chi}^0_1$, the single stop production will give the mono-top
signature \cite{monotop-0,monotop-1,monotop-2,monotop-3,monotop-4,wu} and in Ref.~\cite{wu}
its full-hadronic final states with top tagging technique are studied.
In this study, we focus on the leptonic mono-top channel of the process $pp \to \tilde{t}_1 \tilde{\chi}^-_1$. In contrast with the full hadronic channel, the leptonic channel has no QCD background pollution. Besides, the cut on the leptonic $m_T$ can greatly reduce the $t\bar{t}$ and $W+b$ backgrounds \cite{monotop-2,monotop-3}. We also construct a new variable, which is the open angle of the charged lepton and $b$-jet from the top quark in the stop decay, to reduce the SM backgrounds.

This work is organized as follows. In Sec.~II, we calculate the single stop production at the LHC and stop decays in compressed electroweakino scenarios. In Sec.~III, we perform detailed Monte Carlo simulation for the leptonic mono-top signature from the single stop production at the LHC. Finally, we summarize our conclusions in Sec.~IV.

\section{single production and decays of stop in compressed electroweakino scenario}
In the MSSM, the stop mass matrix in the gauge-eigenstate basis ($\tilde{t}_L$,$\tilde{t}_R$) is given by
  \begin{eqnarray}
  M_{\tilde{t}}^2=
\left(
   \begin{array}{cc}
     m_{\tilde{t}_L}^2 &m_tX_t^{\dag}\\
     m_tX_t& m_{\tilde{t}_R}^2\\
   \end{array}
 \right)
\end{eqnarray}
with
\begin{eqnarray}
&&m_{\tilde{t}_L}^2=m_{\tilde{Q}_{3L}}^2+m_t^2+m_Z^2\left(\frac{1}{2}-\frac{2}{3}\sin^2\theta_W\right)\cos2\beta,\\
&& m_{\tilde{t}_R}^2=m_{\tilde{U}_{3R}}^2+m_t^2+\frac{2}{3}m_Z^2\sin^2\theta_W\cos2\beta,\\
&& X_t = A_t -\mu \cot\beta.
\end{eqnarray}
Here $m_{\tilde{Q}_{3L}}$ and $m_{\tilde{U}_{3R}}$ are the soft-breaking mass parameters for the third generation
left-handed squark doublet $\tilde{Q}_{3L}$ and the right-handed stop $\tilde{U}_{3R}$, respectively.
$A_t$ is the stop soft-breaking trilinear parameter. The generation mixing is neglected here. This hermitian matrix can be diagonalized by
a unitary transformation:
 \begin{eqnarray}
  \left(
    \begin{array}{c}
      \tilde{t}_1 \\
      \tilde{t}_2 \\
    \end{array}
  \right)
  =
\left(
   \begin{array}{cc}
     \cos\theta_{\tilde{t}} &\sin\theta_{\tilde{t}}\\
     -\sin\theta_{\tilde{t}}& \cos\theta_{\tilde{t}}\\
   \end{array}
 \right)
 \left(
    \begin{array}{c}
      \tilde{t}_L \\
      \tilde{t}_R \\
    \end{array}
  \right),
\end{eqnarray}
where $\theta_{\tilde{t}}$ is the mixing angle between left-handed ($\tilde{t}_L$) and right-handed
($\tilde{t}_R$) stops. In the mass eigenstates, the relevant interactions of the stop and electroweakinos
are given by
\begin{eqnarray}
    {\cal L}_{\tilde{t}_1\bar{b}\tilde{\chi}^+_i} &=& \tilde{t}_1
    \bar{b} ( f^{C}_L P_L + f^{C}_R P_R ) \tilde{\chi}^+_i +h.c.~,
    \label{top-t-c} \\
    {\cal L}_{\tilde{t}_1\bar{t}\tilde{\chi}^0_i} &=& \tilde{t}_1
    \bar{t} ( f^{N}_L P_L + f^{N}_R P_R ) \tilde{\chi}^0_i + h.c.~,
    \label{bottom-t-c}
\end{eqnarray}
where $P_{L/R}=(1\mp\gamma_5)/2$,  and
\begin{eqnarray}
    &&f^N_L =
    -\left[ \frac{ g_2 }{\sqrt{2}}N_{i 2}
        + \frac{ g_1}{3\sqrt{2}} N_{i 1}
    \right] \cos\theta_{\tilde{t}} -y_t N_{i 4} \sin\theta_{\tilde{t}}~~~~
\label{left-handed} \\
    &&f^N_R = \frac{2\sqrt{2}}{3} g_1
    N^*_{i 1} \sin\theta_{\tilde{t}}
    - y_t N^*_{i 4} \cos\theta_{\tilde{t}},\\
    &&    f^C_L = y_b U^*_{i 2} \cos\theta_{\tilde{t}},\\
    && f^C_R = - g_2 V_{i 1} \cos\theta_{\tilde{t}}
    + y_t V_{i 2}\sin\theta_{\tilde{t}} ,
\label{right-handed}
\end{eqnarray}
with $y_t=\sqrt{2}m_t/(v\sin\beta)$ and $y_b=\sqrt{2}m_b/(v\cos\beta)$ being the top and bottom Yukawa
couplings, respectively. When $\tan\beta$ is large, the values of $y_b$ can be sizable. The neutralino
and chargino mixing matrices $N_{ij}$, $U_{ij}$, $V_{ij}$ are defined in \cite{mssm-feynrules}. The
compressed electroweakino spectrum,
$m_{\tilde{\chi}^\pm_1}-m_{\tilde{\chi}^0_1} \ll m_{\tilde{\chi}^0_1}$,
can be realized in
two limits:
\begin{itemize}
  \item[(i)] $\mu \ll M_{1,2}$, $V_{11}, U_{11}, N_{11,12,21,22} \sim 0$, $V_{12} \sim \mathop{\rm sgn}(\mu)$,
$U_{12} \sim 1$ and $N_{13,14,23}=-N_{24} \sim 1/\sqrt{2}$. In this limit, the two neutralinos
$\tilde{\chi}^0_{1,2}$ and the chargino $\tilde{\chi}^\pm_1$ are nearly degenerate higgsinos ($\tilde{H}^\pm$).
Such a higgsino LSP scenario may be probed at the high luminosity LHC \cite{giudice-higgisno,wu-higgsino,han-higgsino,park-higgsino,barducci-higgsino}.
  \item[(ii)] $M_2 \ll \mu, M_1$, $V_{11}, U_{11} \sim 1$, $V_{12}, U_{12} \sim 0$, $N_{11,13,14}$, $N_{22,23,24} \sim 0$, and $N_{12,21} \sim 1$. In this case, the lightest neutralino $\tilde{\chi}^0_{1}$ and the lighter chargino
$\tilde{\chi}^\pm_1$ are nearly degenerate winos ($\tilde{W}^\pm$). If the small splitting between
$\tilde{\chi}^\pm_1$ and $\tilde{\chi}^0_{1}$ is not too small, the mono-jet with soft photon events
can be used to detect this wino LSP scenario at the LHC \cite{braman-wino,han-wino,ismail-wino}.
\end{itemize}

We evaluate the mass spectrum and branching ratios of all sparticles with \textsf{SUSY-HIT} \cite{SUSYHIT}.
We use \textsf{MadGraph5\_aMC@NLO} \cite{Madgraph} to calculate the leading order cross section of the single
stop production. The NNPDF23LO1 \cite{nnpdf} parton distribution functions are chosen for our calculations.
The renormalization and factorization scales are set as the default value. We include the NLO-QCD correction by applying a $K$-factor of 1.3 \cite{single-stop-1,wu} to the cross section of the single stop production. It should be noted that
the single stop production not only relies on the nature of the electroweakinos, but also is affected by the
polarized states of the stop. To demonstrate this, we consider two cases: the left-handed stop $\tilde{t}_L$ by taking
$m_{\tilde{U}_{3R}}=2$ TeV to decouple the right-handed component, and the right-handed stop $\tilde{t}_R$
by taking $m_{\tilde{Q}_{3L}}=2$ TeV to decouple the left-handed component.

\begin{figure}[ht!]
   \centering
  \includegraphics[height=5in,width=6in]{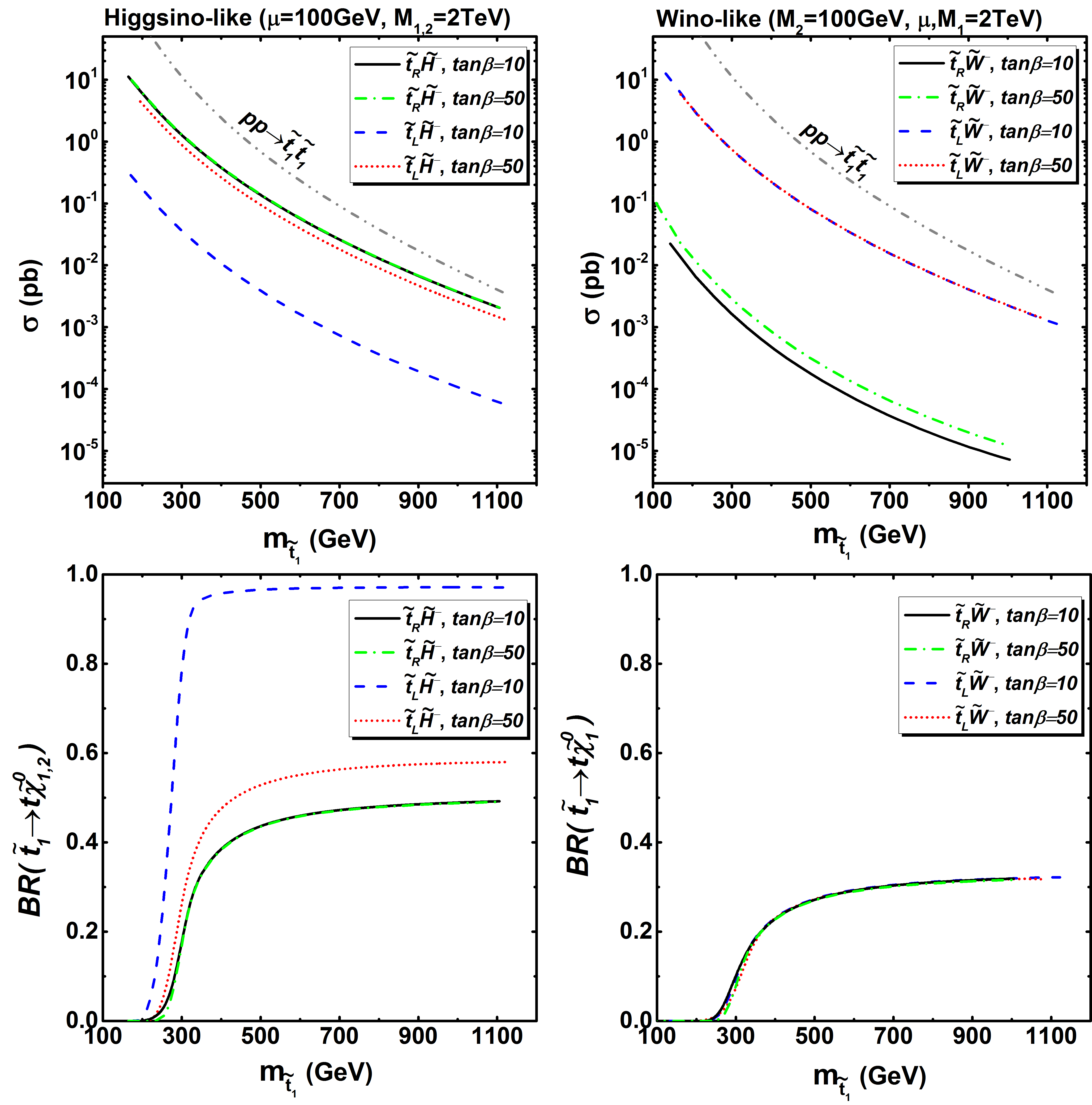}
\vspace{-0.5cm}
\caption{Cross sections of the associated production of stop and chargino at 14 TeV LHC
(\emph{upper panel}), and the stop decay branching ratios (\emph{lower panel}) in two compressed electroweakino scenarios, where $\tan\beta=10$ and 50. The left (right) two figures are for a higgsino-like (wino-like) chargino $\tilde\chi^\pm_1$. }
  \label{fig:cross}
\end{figure}
In the upper panels of Fig.~\ref{fig:cross}, we show the cross sections of the associated production of stop and chargino at 14 TeV LHC for four different final states:
$\tilde{t}_L \tilde{H}^-$, $\tilde{t}_R \tilde{H}^-$, $\tilde{t}_L \tilde{W}^-$ and
$\tilde{t}_R \tilde{W}^-$. The contributions of the conjugate processes are included.
For a higgsino-like chargino, we can see that the cross section of $\tilde{t}_R\tilde{H}^-$ production is larger than that of $\tilde{t}_L\tilde{H}^-$ production and almost independent of $\tan\beta$. It can reach about 3 pb at $m_{\tilde{t}_1}=200$ GeV. While the cross section of $\tilde{t}_L\tilde{H}^-$ strongly depends on the value of $\tan\beta$, since the coupling of the left-handed stop with $\tilde{\chi}^\pm_{1}$ is dominated by the bottom Yukawa coupling $y_b$ and can be enhanced by a large $\tan\beta$. For a wino-like chargino, the cross section of
$\tilde{t}_L\tilde{W}^-$ is always much larger than that of $\tilde{t}_R\tilde{W}^-$ because of the gauge interactions.

In the lower panel of Fig.~\ref{fig:cos}, we present the branching ratios of stop decaying to the top quark and neutralinos. For higgsino case, it can be seen that a left-handed stop $\tilde{t}_L$ dominantly decays to $t \tilde{\chi}^0_{1,2}$ at $\tan\beta=10$. The reason is that the decay width of $b \tilde{\chi}^+_{1}$ is proportional to $y_b$ and is suppressed for a small $\tan\beta$. If the stop is right-handed $\tilde{t}_R$, its couplings with $\tilde{\chi}^0_{1,2}$ and $\tilde{\chi}^\pm_{1}$ are proportional to $y_t$, and the branching ratios of $\tilde{t}_{R} \to t \tilde{\chi}^{0}_{1,2}$ and
$\tilde{t}_R \to b \tilde{\chi}^+_1$ are about 50\% and 50\%, respectively. For the wino case, both $\tilde{t}_L$ and $\tilde{t}_R$ decay to $t\tilde{\chi}^0_1$ with the same branching ratio.

Besides, it can be seen that the cross section of stop pair production $\sigma(\tilde{t}\tilde{t}^*)$ is about one order of magnitude larger than that of single stop production if stop mass is less than 300 GeV. With the increase of stop mass, the cross section of stop pair production decreases more rapidly than the single stop production due to the suppression of phase space. For example, when stop mass is 700 GeV, the ratio of $\sigma(\tilde{t}\tilde{t}^*)/\sigma(\tilde{t}_R \tilde{H}^-)$ is about four. Considering the stop decay branching ratios, we find that the number of events of $\tilde{t}_1\tilde{t}^*_1(\to \bar{t} \tilde{\chi}^0_{1,2})$ production is still about two times larger than that of $\tilde{t}_R(\to t \tilde{\chi}^0_{1,2})\tilde{H}^-$ production. While the expected number of events of $\tilde{t}_L(\to t \tilde{\chi}^0_{1,2})\tilde{H}^-$ and $\tilde{t}_{L,R}(\to t \tilde{\chi}^0_1)\tilde{W}^-$ productions are less than that of $\tilde{t}_R(\to t \tilde{\chi}^0_{1,2})\tilde{H}^-$. In the following, we will use $\tilde{t}_R(\to t \tilde{\chi}^0_{1,2})\tilde{H}^-$ production as an example to investigate the observability of the single stop production at the LHC.

\section{leptonic mono-top signature from single stop production at the LHC}
Since $\tilde{\chi}^\pm_1$ and $\tilde{\chi}^0_{1,2}$ are the nearly degenerate higgsinos in our
considered scenario, the mass splitting between them is small so that $\tilde{\chi}^\pm_1$
and $\tilde{\chi}^0_2$ appear as missing transverse energy at the LHC. This leads to
the mono-top signature for the single stop production at the LHC, which is
\begin{equation}
pp \to \tilde{t}_1(\to t\tilde{\chi}^0_{1,2}) \tilde{\chi}^-_1 \to t +\slashed E_T,
\end{equation}
In our simulation, we focus on the leptonic mono-top channel. In contrast with the full hadronic final states, the problematic QCD multijet background can be safely neglected in this leptonic channel. We use \textsf{MadGraph5\_aMC@NLO} \cite{Madgraph} to generate the parton level events.
Then, we perform the parton shower and hadronization by \textsf{Pythia} \cite{pythia}.
The jets are clustered by the anti-$k_t$ algorithm with a cone radius $\Delta R = 0.4 $ \cite{antikt}.
We implement the detector effects with \textsf{Delphes} \cite{delphes}.

The SM backgrounds are dominated by the following processes:
\begin{itemize}
  \item The largest SM backgrounds are the semi- and di-leptonic $t\bar{t}$ productions,
where the missing lepton and the limited jet energy resolution will lead to relatively
large missing $E_T$. The leading order cross section of $t\bar{t}$ production is
normalized to its approximate next-to-next-to-leading-order value
$\sigma_{t\bar{t}}^{\mathrm{NNLOapprox}}=920$ pb \cite{tt}.
 \item The subdominant background is the single top production, which is irreducible, up to a jet
that could come from ISR. We include three production modes $tj$, $tb$ and $tW$ in our simulation.
\end{itemize}
There are other possible SM backgrounds, such as $W + {\mathrm jets}$ and the diboson production $VV$. But for $W + {\mathrm jets}$,
the mistag rate of a light jet as a $b$-jet in current ATLAS and CMS analyses is of the
order of $10^{-2}$ and $10^{-3}$, depending on the working point of the $b$-tagging algorithm.
The acceptance of this background after cuts is found to be negligibly small. On the other hand,
$VV$ backgrounds can also be neglected because of their small cross sections and the difficulty
of faking a $b\ell \slashed{E}_T$ final state in $WW$, $WZ$ and $ZZ$ backgrounds.

\begin{figure}[ht!]
   \centering
  \includegraphics[width=3.in,height=3.5in]{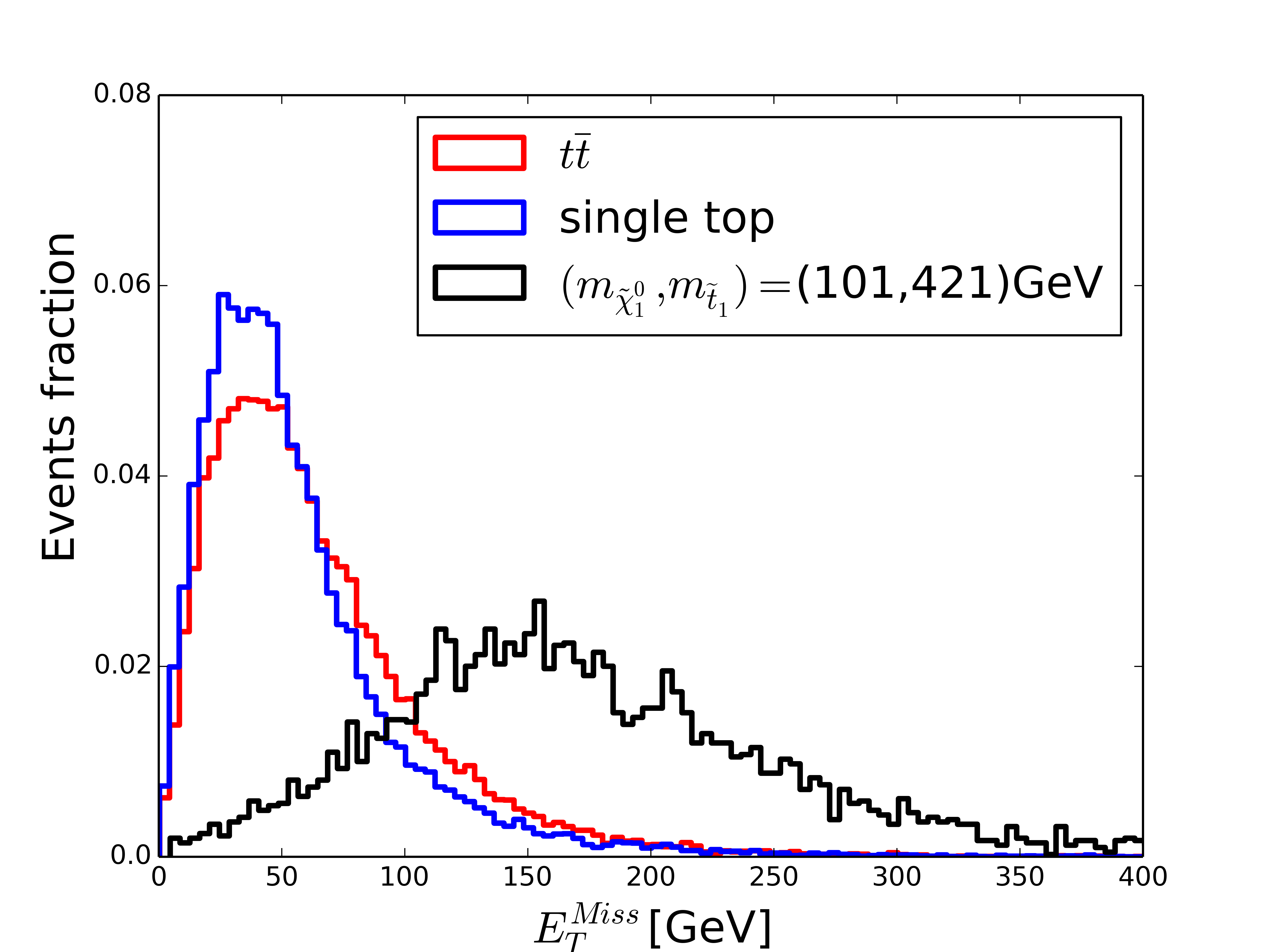}\hspace{-0.5cm}
  \includegraphics[width=3.in,height=3.5in]{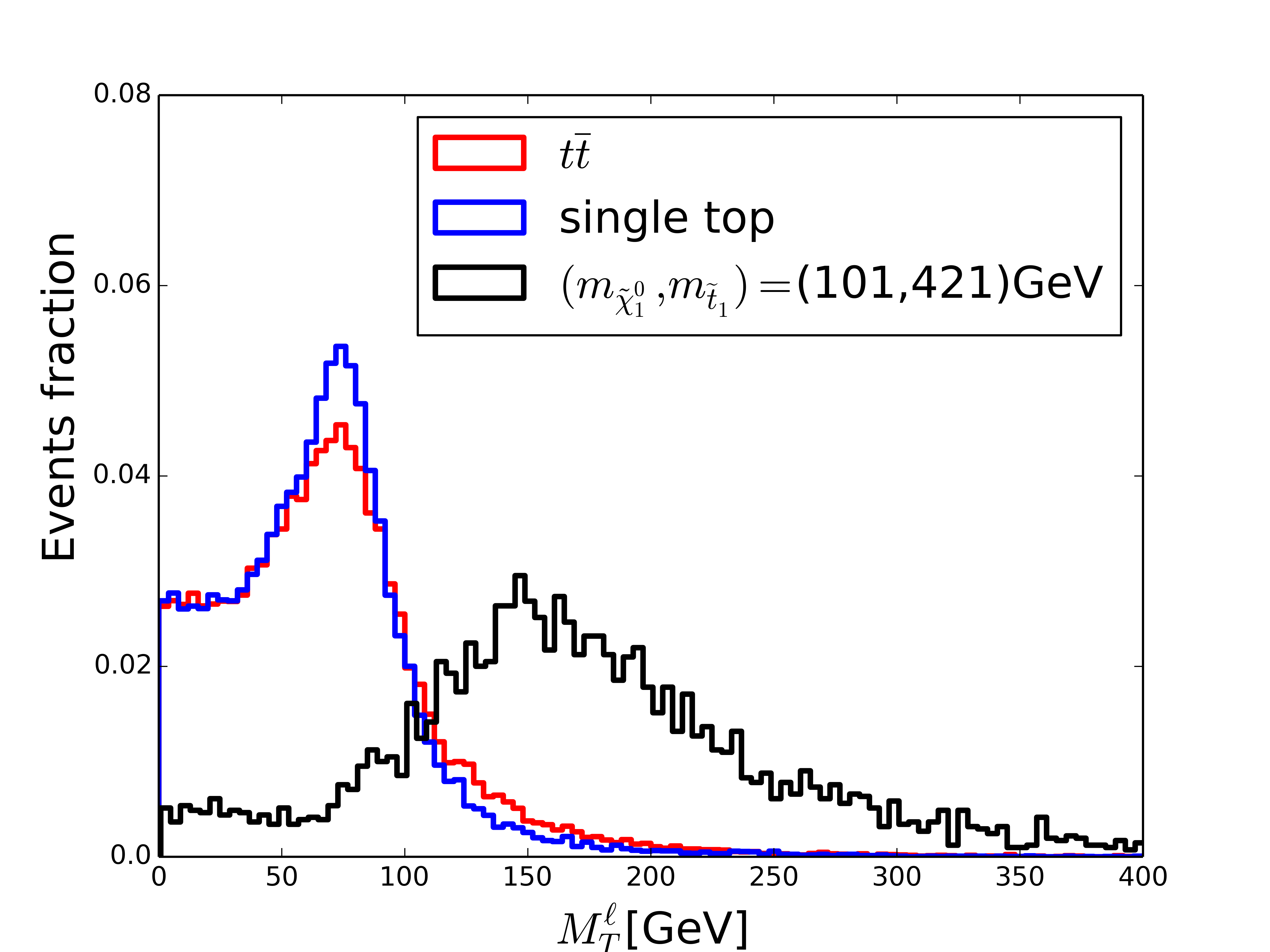}
\vspace{-0.3cm}
  \caption{Distributions of transverse missing energy $\slashed E_{T}$ and the transverse mass
of the lepton plus missing energy system $M_T^\ell$. The signal benchmark point is for
$m_{\tilde{\chi}^0_1}=101$ GeV and $m_{\widetilde{t}_1}=421$ GeV.}
  \label{fig:mtw}
\end{figure}
In Fig.~\ref{fig:mtw}, we present the distributions of the transverse missing energy $\slashed E_{T}$ and
the transverse mass of the lepton plus missing energy system $M_T^\ell$. It is clear that the backgrounds
and the signal can be discriminated by $\slashed E_{T}$. Most events of the backgrounds are distributed
in the region of $\slashed E_{T} \lesssim 150~ \rm GeV$. However, the signal has much more events than
backgrounds in the region of $\slashed E_{T} \gtrsim 150~ \rm GeV$, due to the extra missing energy from
the massive LSP. Besides, the variable $M_T^\ell$ can well separate the backgrounds and signal because
it has an end-point at the mass of the lepton's parent particle, $M^\ell_{T}|_{\mathrm max}=M$~\cite{pdg}.
All the main backgrounds contain a $W$ boson and a unique source of missing energy, the neutrino,
coming from its decay. So the backgrounds have endpoint around $M_W$ in the $M_T^\ell$ distributions.
But the signal has a larger value of $M^\ell_{T}$. A cut on $M^\ell_{T} \ge 80 ~ \rm GeV$ will greatly
reduce the backgrounds while keep most of the signal.

\begin{figure}[ht!]
   \centering
  \includegraphics[width=4in,height=3.5in]{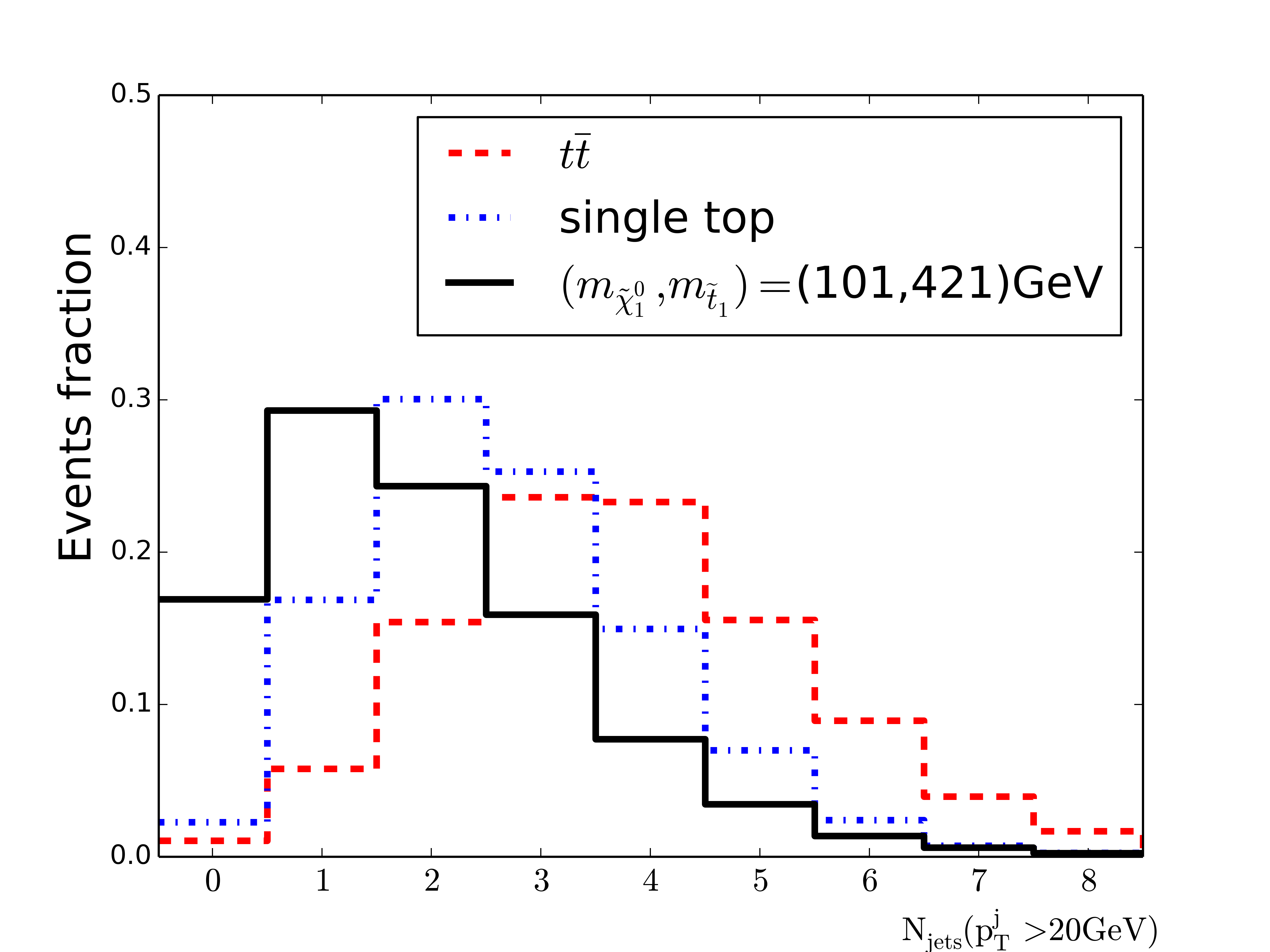}
\vspace{-0.5cm}
  \caption{Same as Fig.~\ref{fig:mtw}, but for the distribution of jet multiplicity.}
  \label{fig:jet}
\end{figure}
In Fig.~\ref{fig:jet}, we show the jet multiplicity ($N_{jets}$) distributions of the signal and backgrounds. We can see that most of events of $t\bar{t}$ and single top backgrounds have larger $N_{jets}$ than the single stop process. To suppress the backgrounds, we will veto the second hard jet in our event selection.

\begin{figure}[ht!]
   \centering
  \includegraphics[width=4in,height=3.5in]{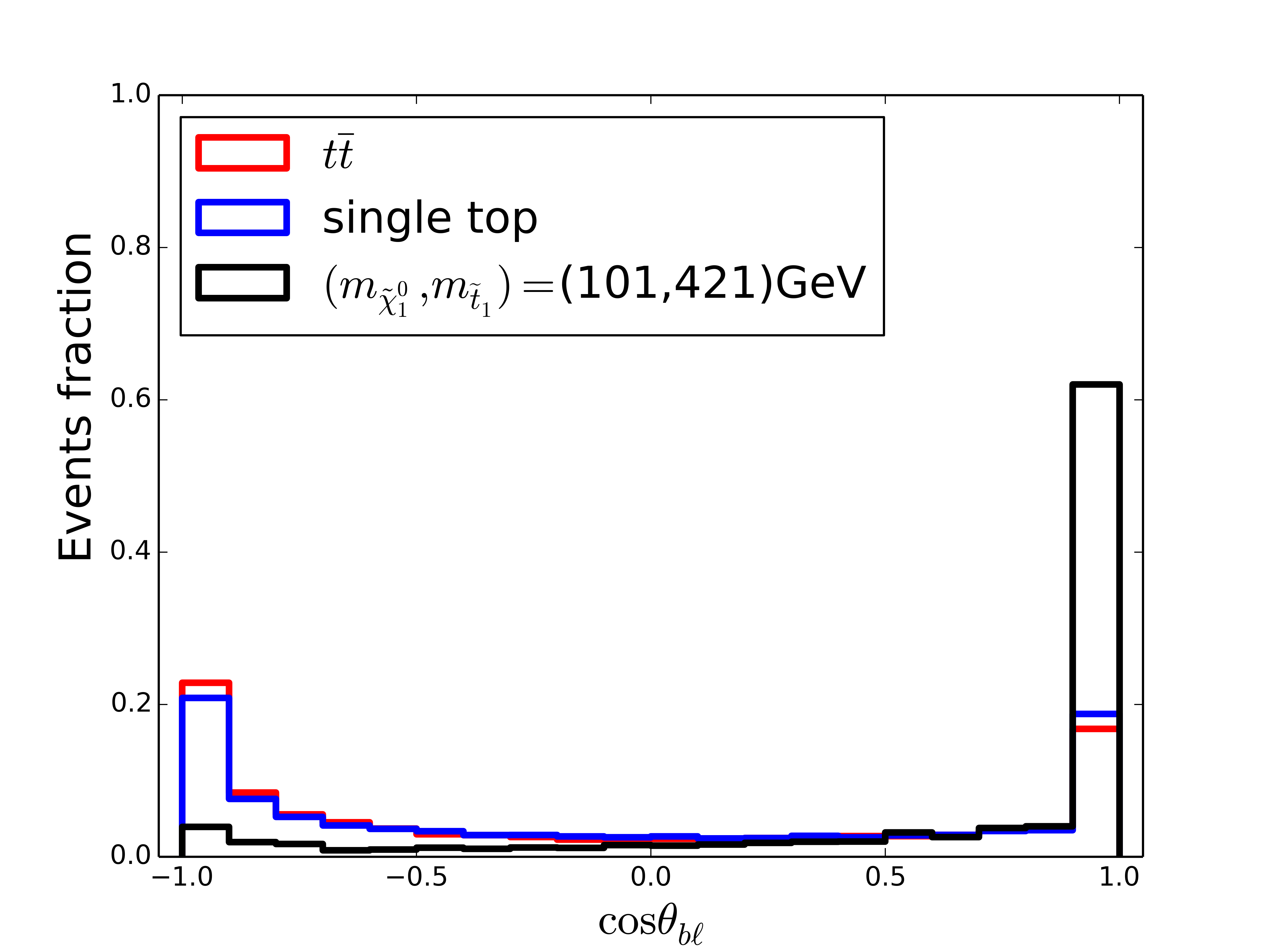}
\vspace{-0.5cm}
  \caption{Same as Fig.~\ref{fig:mtw}, but for the distribution of the opening angle $\cos\theta_{b\ell}$
between the charged lepton and the $b$-jet in the lab-frame.}
  \label{fig:cos}
\end{figure}
Another interesting observable is the opening angle $\theta_{b\ell}$ between the charged
lepton and the $b$-jet in the lab-frame. After requiring exactly one lepton and one $b$-jet, we display the
distribution of $\cos\theta_{b\ell}$ in Fig.~\ref{fig:cos}. We can see that most of the signal events fall
in the region of $\cos\theta_{b\ell}>0$, while the backgrounds have more events in the region of
$\cos\theta_{b\ell}<0$. This is because the the charged lepton and the $b$-quark from top quark in the stop
decays are boosted so that they tend to move in the same direction when the mass splitting between
$\tilde{t}_1$ and $\tilde{\chi}^0_1$ is large. Thus, the requirement of a large $\cos\theta_{b\ell}$ can
further reduce backgrounds.

\begin{figure}[ht!]
   \centering
  \includegraphics[width=4.5in,height=3.5in]{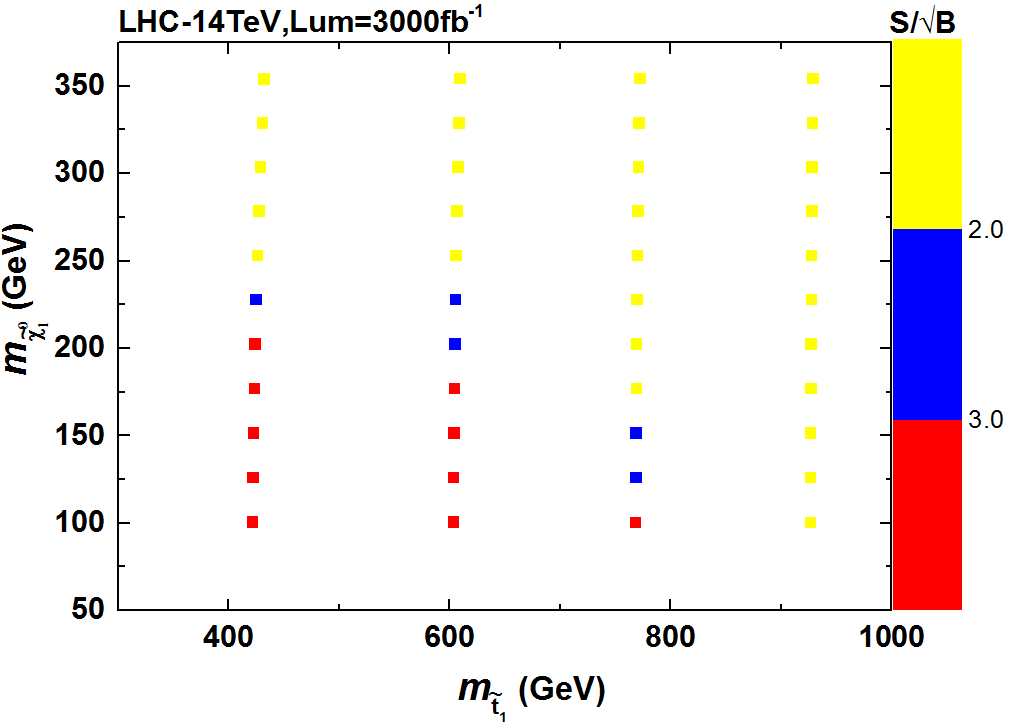}
\vspace{-.5cm}
  \caption{The statistical significance $S/\sqrt{B}$ on the plane of $m_{\tilde{t}_1}$ versus $m_{\tilde{\chi}^0_1}$ at 14 TeV LHC with $\mathcal{L} = 3000 $ fb$^{-1}$.}
  \label{fig:cover}
\end{figure}

The detailed analysis strategies are the followings:
\begin{itemize}
  \item We require exact one hard lepton with $p_{T} (\ell) >30$ GeV and  $|\eta_\ell|<2.5$.
  \item We require exact one $b$-jet with $p_T(b)>75$ GeV and $|\eta_b|<2.5$ and veto extra jets with $p_T(j) > 20$ GeV to suppress the $t\bar{t}$ background.
  \item We define eight signal regions according to $\slashed E_T$ cuts: 150, 175, 195, 200, 205, 225, 250 and 275 GeV. It is worth noting at this point that cuts in $M_T$ end up having little correlation with cuts
in $\slashed E_{T}$, as it will be shown in the cut-flow tables below.
  \item We require $M^\ell_T>175$ GeV and $\cos\theta_{b\ell}>0.85$ to suppress top pair and single stop backgrounds.
\end{itemize}
Finally, we use the signal region with highest $S/\sqrt{B}$ to show our results in Fig.~\ref{fig:cover}.

\begin{table}[ht!]
\caption{A cut flow analysis of the cross sections of the backgrounds and signal at 14 TeV LHC, where the cross sections are in unit of fb. The significance $S/\sqrt{B}$ is calculated assuming $3000 fb^{-1}$ of integrated luminosity. The benchmark point is $(m_{\tilde{\chi}_1^0},m_{\tilde{t}_1})=(101,421)$ GeV. }
\footnotesize\begin{tabular}{|c|c|c|c|c|c|c|}
  \hline
  cut & 1 lepton & 1 $b$-jet & jet veto & $M_T^{\ell}>$ 175 &$\slashed E_T >$150 &$\cos \theta_{b\ell}$ \\
    &$p_T^\ell>30GeV,|\eta^\ell|<2.5$&  $p_T^b>75GeV,|\eta^b|<2.5$&$p_T(j) >20$GeV&[GeV] &[GeV]&$>$0.85\\
  \hline
  $t\bar{t}$ &233465.16& 77973.38 & 796.20 &62.95 &26.2472&11.75\\
  \hline
  $t+j/b/W$ &44891.80&8411.10&189.24&9.45&3.47&1.64\\
  \hline
  signal & 24.88& 9.482& 1.40&1.03&0.90&0.77 \\
  \hline
  S/B(\%)&&&&1.43&3.02&5.75\\
  $S/\sqrt{B}$&&&&6.67&9.04&11.53\\
  \hline
  \end{tabular}
\label{tab:cutflow}
\end{table}
In Table~\ref{tab:cutflow}, we present a cut flow of cross sections for the signal and backgrounds at the
14 TeV LHC. The benchmark point is $m_{\tilde{\chi}^0_1}=101$ GeV and $m_{\widetilde{t}_1}=421$ GeV.
We can see that the $t\bar{t}$ production is the largest SM background. The requirement of exact one
$b$-jet with $p_T^b > 75$ GeV can reduce the backgrounds by about 60\%. The jet-veto for the second hard
jet can significantly reduce $t\bar{t}$ background by almost two orders of magnitude. The cuts of $M^\ell_T > 175$ GeV
and $\slashed E_T > 150$ GeV can further remove the backgrounds by one order of magnitude. It should be noted
that $\cos\theta_{b\ell}>0.85$ can help to suppress backgrounds by half and improve the value of $S/B$.

In Fig.~\ref{fig:cover}, we plot the dependence of the signal significance $S/\sqrt{B}$ on $m_{\tilde{\chi}^0_1}$
and $m_{\widetilde{t}_1}$ for the 14 TeV LHC with a luminosity $\mathcal{L} = 3000$ fb$^{-1}$. From this figure
we can see that the significance drops with the increase of $m_{\tilde{\chi}^0_1}$ and $m_{\widetilde{t}_1}$
because of the reduction of the cross section. We find that the parameter range
$100~{\rm GeV} \leq m_{\tilde{\chi}^0_1} \leq 150~{\rm GeV}$ and
$m_{\tilde{t}_1} \leq 760~{\rm GeV}$ can be covered at
$\geq 2\sigma$ level with $S/B > 3\%$ at the HL-LHC, which is moderately better than the hadronic stop channel \cite{wu}.

\begin{figure}[ht!]
   \centering
  \includegraphics[width=4.5in,height=3.5in]{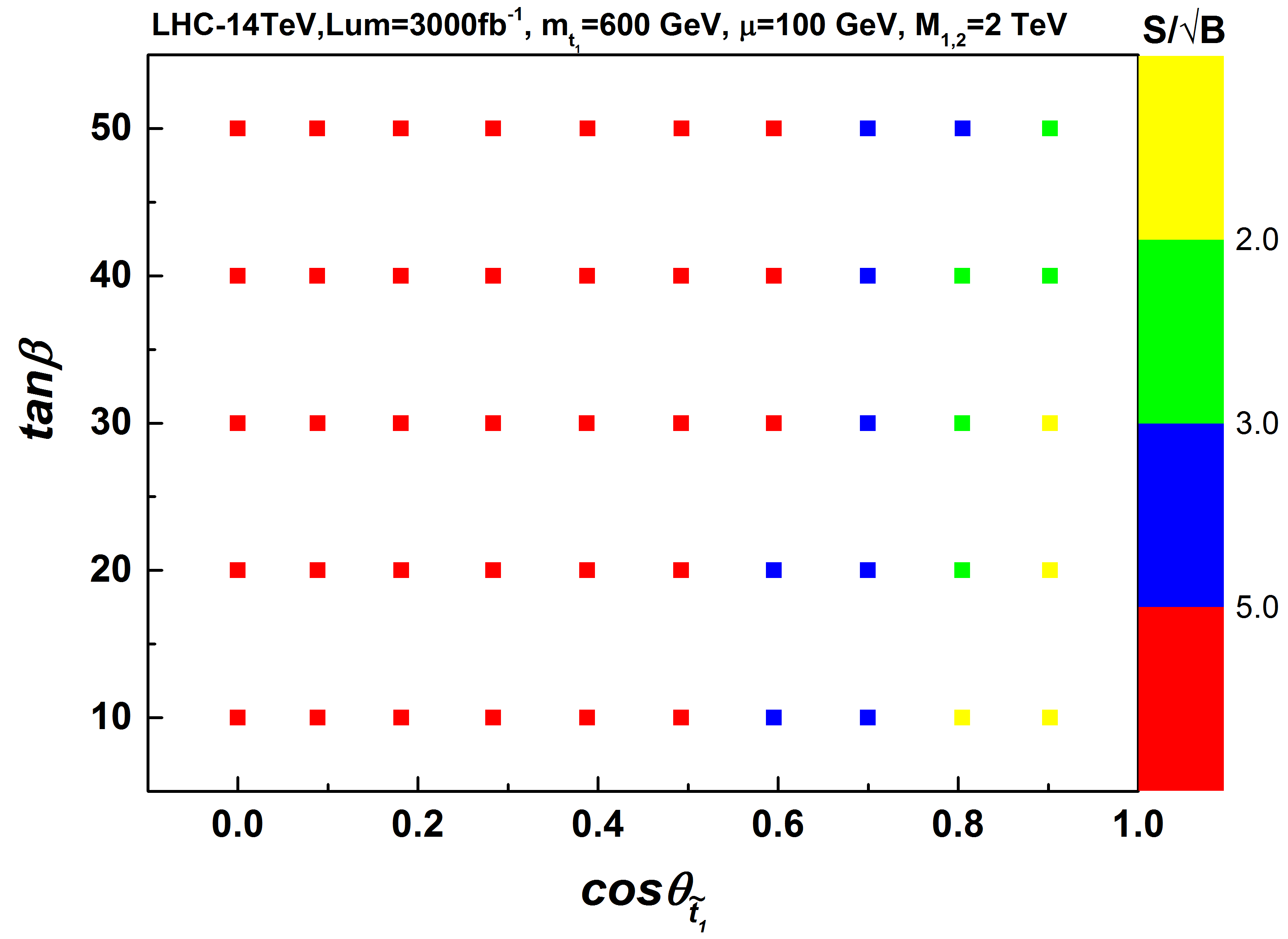}
\vspace{-.5cm}
  \caption{The statistical significance $S/\sqrt{B}$ for a benchmark point $m_{\tilde{t}_1}=600$ GeV, $\mu=100$ GeV and $M_{1,2}=2$ TeV on the plane of $\cos\theta_{\tilde{t}_1}$ and $\tan\beta$ at 14 TeV LHC with $\mathcal{L} = 3000 $ fb$^{-1}$.}
  \label{fig:signal}
\end{figure}
Given a discovery of the stop and a measurement of the single stop production cross
section, we examine the discriminating power of single stop production with regard to the electroweak properties of the stop. In Fig.~\ref{fig:signal}, we show the statistical significance $S/\sqrt{B}$ of the process $pp \to \tilde{t}_1\tilde{\chi}^-_1$ for a benchmark point $m_{\tilde{t}_1}=600$ GeV, $\mu=100$ GeV and $M_{1,2}=2$ TeV on the plane of stop mixing angle $\cos\theta_{\tilde{t}_1}$ and $\tan\beta$ at 14 TeV LHC with $\mathcal{L} = 3000$ fb$^{-1}$. We can see that the mixing angle $\cos\theta_{\tilde{t}_1}\lesssim 0.5$ ( right-handed-like stop) can be probed above $5\sigma$ level. While for $\cos\theta_{\tilde{t}_1} \gtrsim 0.5$ (left-handed-like stop), the significance $S/\sqrt{B}$ depends on the value of $\tan\beta$. This is because the cross section of $\tilde{t}_L\tilde{H}^-$ production is sensitive to $\tan\beta$ (c.f. Fig.~\ref{fig:cross}). When $\cos\theta_{\tilde{t}_1}>0.7$, the significance $S/\sqrt{B}$ is be less than $3\sigma$. On the other hand, if $\tilde{\chi}^-_1$ is wino-like, we can expect that the large $\cos\theta_{\tilde{t}_1}$ region will have larger $S/\sqrt{B}$ than the small $\cos\theta_{\tilde{t}_1}$ region at the HL-LHC since the cross section of $\tilde{t}_L\tilde{W}^-$ production is much larger than that of $\tilde{t}_R\tilde{W}^-$ production (c.f. Fig.~\ref{fig:cross}).

\section{CONCLUSION}
In this work we explored the observability of the associated production of stop and chargino in the compressed electroweakino scenario at 14 TeV LHC. Due to the small mass splitting between $\tilde\chi^0_1$ and $\tilde{\chi}^-_1$, such a production can lead to the mono-top
signature via stop decay $\tilde{t}_1 \to t \tilde{\chi}^0_1$. We analyze the leptonic mono-top channel $pp \to \tilde{t}_1\tilde{\chi}^-_1 \to b\ell+\slashed E_T$, and construct a lab-frame observable $\cos\theta_{b\ell}$ from the top quark in
the stop decay to reduce the SM backgrounds. We found that the stop mass can be probed up to 760 GeV at $2\sigma$ level through the single stop production at 14 TeV LHC with $\mathcal{L} = 3000$ fb$^{-1}$. We also find that the stop mixing angle can also be determined from the single stop production assuming a measurement of the single stop production cross section at HL-LHC.

\section*{ACKNOWLEDGMENTS}
G. H. Duan thanks Yang Zhang for helpful discussions.
This work is partly supported by the Australian Research Council,
by the National Natural Science Foundation of China (NNSFC)
under grants Nos. 11105124, 11105125, 11275057, 11305049, 11375001, 11405047, 11135003, 11275245,
by the CAS Center for Excellence in Particle Physics (CCEPP)
and by the CAS Key Research Program of Frontier Sciences.

\end{document}